\documentstyle[aps,multicol]{revtex}

\renewcommand{\narrowtext}{\begin{multicols}{2}\global\columnwidth20.5pc}
\renewcommand{\widetext}{\end{multicols}\global\columnwidth42.5pc}

\input{epsf.tex}

\begin{document}

\draft

\tighten

\title{Nature of Spin Excitations in Two-dimensional 
Mott Insulators: Undoped Cuprates and Other Materials} 
\author{Chang-Ming Ho$^{a,}$$^b$, V. N. Muthukumar$^c$, Masao Ogata$^d$ and
P. W. Anderson$^c$}
\address{$^a$ Department of Applied Physics, University of Tokyo, 7-3-1 Hongo, 
Bunkyo-ku, Tokyo 113-8656, Japan\\
$^b$ Physics Division, National Center for Theoretical Sciences, 
P.O.Box 2-131, Hsinchu, Taiwan 300 $\dagger$\\
$^c$ Joseph Henry Laboratories of Physics, Princeton University,
Princeton, NJ 08544\\
$^d$ Department of Physics, University of Tokyo, 7-3-1 Hongo, Bunkyo-ku, 
Tokyo 113-0033, Japan\\
}
\date{September, 2000. {\em Published in} Phys. Rev. Lett. 86, 1626 (2001)}
\maketitle

\begin{abstract}
We investigate the excitation spectrum of a two-dimensional 
resonating valence bond 
(RVB) state. Treating the $\pi$-flux phase with antiferromagnetic
correlations as a variational ground state, we recover the long
wavelength magnon as an ``RVB exciton''. However, we find that this
excitation does {\em not} exhaust the entire spectral weight and the 
high energy 
spectrum is dominated by fermionic excitations. 
The latter can be observed directly by inelastic neutron scattering and
we predict their characteristic energy scales along different high
symmetry directions in the magnetic Brillouin zone. We also interpret
experimental results on two magnon Raman scattering and mid-infrared
absorption within this scenario.
\end{abstract}


\narrowtext

The undoped high $T_{c}$ cuprates, such as $La_2CuO_4$ {\it etc.}, 
\cite{rmp} and compounds like $Cu$($DCO_{2}$)$_{2} \cdot 4D_{2}O$ (CFTD) 
\cite{cftd,clarke} are ideal 
realizations of two-dimensional (2D) spin 1/2 
Heisenberg antiferromagnets on the square-lattice composed of $Cu$ ions and, 
consequently, are interesting systems to study. The antiferromagnetism that
is observed in the undoped cuprates plays a crucial role in some theories
of the metallic state of doped Mott insulators. For these reasons, it is 
important to understand the nature of magnetic excitations in these materials.
A direct way to probe the spectrum of magnetic excitations is 
inelastic neutron scattering (INS). The presence of long 
range antiferromagnetic order below the N\'{e}el
temperature was established in the insulating cuprates by neutron
scattering measurements \cite{rmp}. Well defined peaks, identified as magnons, 
have been observed to disperse along 
the magnetic zone boundary in different compounds 
\cite{clarke,hayden_91}. Since then, a spin wave dispersion throughout the
entire Brillouin zone in $La_{2}CuO_{4}$, has been extracted from 
INS measurements \cite{coldea_00}.
The antiferromagnetic correlation length diverges exponentially as the
temperature is lowered, indicative of long range order at 
$T=0$ \cite{rmp,cftd}. 
This is in agreement with the analysis of
the 2D antiferromagnetic Heisenberg model, with superexchange 
coupling $J$ between adjacent spins,
in terms of the 
quantum non-linear $\sigma$ model \cite{ch2n2}, and other 
approaches based 
on spin waves \cite{pqscha}. 
These results, concerning essentially the low energy regime, 
are construed as 
evidence favoring the hypothesis that the
excitation spectrum in these materials can be described adequately by
renormalized spin wave theory \cite{manousakis_91}.

On the other hand, optical probes such as Raman and mid-infrared
absorption spectroscopy highlight the inadequacy of the spin
wave hypothesis. Typically, the positions of primary peaks observed in
these experiments coincide reasonably well with calculations based on
spin wave theory, but the observed line shapes cannot be explained, owing
to the presence of finite spectral weight at high energies. For
instance, the $B_{1g}$ shift in Raman scattering experiments spans a 
rather broad range of energies in insulating cuprates \cite{sugai}. 
Recent measurements on insulating $Sr_2CuO_2Cl_2$ as well as in 
$YBa_2Cu_3O_{6.1}$
show very broad line shapes, spectral weight at higher energies, 
and a broad feature around $4J$ \cite{blumberg_96}.
Optical absorption spectra 
show similar features.
Experiments on insulating cuprates \cite{perkins,grueninger} 
show a primary peak identified with
bimagnon plus phonon absorption \cite{lorenzana_95}. But as in Raman
scattering, the spectrum has a long tail extending to $8000$ $cm^{-1}$ in
$La_{2}CuO_{4}$ and $6000$ $cm^{-1}$ in $YBa_2Cu_3O_{6}$. 
By analyzing the lineshape 
and the high energy spectra, Gr\"{u}ninger and coworkers have presented 
strong experimental evidence against the spin wave hypothesis 
for the insulating cuprates \cite{grueninger}. 

The two classes of experiments described in the
preceding paragraphs do not necessarily contradict each other.
Though the dispersing peaks observed in INS measurements can be interpreted 
as spin waves, an analysis of the {\em
spectral weight} shows that long range order and spin waves can account
only for about 50\% of the observed spectrum \cite{swt}. 
These results indicate the
presence of excitations beyond the one magnon mode. Thus, 
taken in conjunction, INS and
optical spectroscopy suggest two possible theoretical approaches. 
One is to start from spin wave theory, and look for consistent
explanations for the experimental results outlined above. This is not an
easy task, as spin wave theory is an effective theory of long
wavelength excitations. The other approach is to postulate a new set of
excitations that offers a natural explanation of the high energy 
spectra, and describing the magnons in terms of this new basis.
It is the latter approach that we shall pursue.

In this paper, we consider the flux-phase 
resonating valence bond (RVB) theory, treated within the random 
phase approximation (RPA). Gapless spin wave excitations are 
recovered as a Goldstone mode. Besides this, the RVB theory 
also makes specific predictions about novel excitations- spin 1/2 flux 
fermions or spinons, and their direct observation. 
We predict the characteristic energy scales of these excitations along
various high symmetry directions and show how their presence can
naturally account for the spectral weight seen at high energies in
optical probes.

Our starting point is the $\pi$-flux state with staggered magnetization
as a variational parameter \cite{hsu_90}. For the case of the 2D Heisenberg
antiferromagnet, the above state yields the best variational energy when 
the Gutzwiller constraint is enforced exactly \cite{himeda_99}.
We assume that the excited states of the flux fermion spectrum are also good 
variational states for the spin excitations of 2D Mott insulators. 
The $\pi$-flux state with staggered magnetization $m$ is obtained in the 
mean-field approximation of the Hamiltonian  
\begin{equation}
{\cal H}=-{\frac{J_{\rm eff}}{2}}\sum_{\langle ij \rangle, \sigma} 
e^{i{\Phi}_{\Box}} 
(f^{\dagger}_{i\sigma}f_{j\sigma}+ h.c.)+
V\sum_{i} n_{i\uparrow} n_{i\downarrow}~~,
\label{hamiltonian}
\end{equation}
which is an extension of the RVB mean field theory \cite{flux}.
The lattice fermions pick up a phase 
${\Phi}_{\Box}$=$\pi$ on hopping around an elementary 
plaquette, and also experience an on-site 
potential $V$, which is included to induce a spin density wave
(SDW) coexisting with the flux order. The excitation spectrum is 
given by $E_{\bf k}$=$J_{\rm eff}\sqrt{\cos^{2}k_{x}+\cos^{2}k_{y}
+(m/2)^{2}}$ where $m$ is determined self-consistently as 
$V^{-1}$=$N^{-1}\sum^{\prime} E_{\bf k}^{-1}$, $N$ being the number of 
lattice sites and the summation being over momenta in the magnetic 
Brillouin zone (MBZ).  

The above Hamiltonian, with $J_{\rm eff} \equiv J$, was first 
considered by Hsu \cite{hsu_90}.
Invoking a Gutzwiller approximation, Hsu determined the projected
variational energy and the sublattice magnetization as a function of the
SDW mass parameter $m$, induced by the potential $V$.
The analysis of the excitation spectrum, however, is complicated by the
Gutzwiller approximation. Owing to this difficulty, 
Hsu could only obtain 
the poles of the particle-hole Green's function 
($S_z = \pm 1$ excitations) and not the complete dynamical spin
susceptibility. To obtain the latter, we find it convenient to follow
an equivalent approach, proposed first by Laughlin \cite{laughlin_95}.
In this scheme, 
a spinon pair is created as a projected particle-hole excitation
and the spinons constituting the pair experience an on-site repulsion,
favoring the formation of an SDW. This interaction also
enhances the exchange integral $J$ to $J_{\rm eff}$. The quantities
$J_{\rm eff}$ and $m$ are determined self consistently to be $1.5J$ and
$0.5$, respectively \cite{laughlin_95}.
The Gutzwiller approximation technique employed by Hsu gives similar
results.

We now obtain the dynamical spin susceptibility from
Eq.(\ref{hamiltonian}) using RPA. 
The calculation parallels that of Schrieffer {\em et al.},
on the SDW instability of the 2D Hubbard model \cite{swz_89}. 
We obtain, for the 
(transverse) spin susceptibility, $S({\bf q},\omega) = -{\rm Im}
\chi^{+-}({\bf q},\omega)$, 
where 
\begin{equation}
\chi^{+-}({\bf q},\omega) = {\chi^{+-}_0({\bf q},\omega) \over 
		     1 - V \chi^{+-}_0({\bf q},\omega)}~~,
\label{chi-rpa}
\end{equation}
and
\begin{eqnarray*}
\lefteqn{
\chi^{+-}_0({\bf q},\omega)=-\frac{1}{2N} 
{\sum_{\bf k}}^{\prime} }\nonumber \\
                          & & [1-{\cos k_x\cos (k_x+q_x)+\cos
		       k_y\cos(k_y+q_y)-(m/2)^2 \over E_{\bf k}E_{\bf
		       k+q}}] \nonumber \\
                          & & [{1 \over \omega-(E_{\bf k}+E_{\bf
		       k+q})+i\delta}-{1 \over \omega+(E_{\bf k}+E_{\bf
		       k+q})-i\delta}]~~.
\end{eqnarray*}
Here, $\chi^{+-}_0$ is defined as the time-ordered product 
${\langle TS^{+}S^{-} \rangle}$ with respect to the SDW ground state. 

\begin{figure}
\epsfxsize=3.6in
\centerline{\epsffile{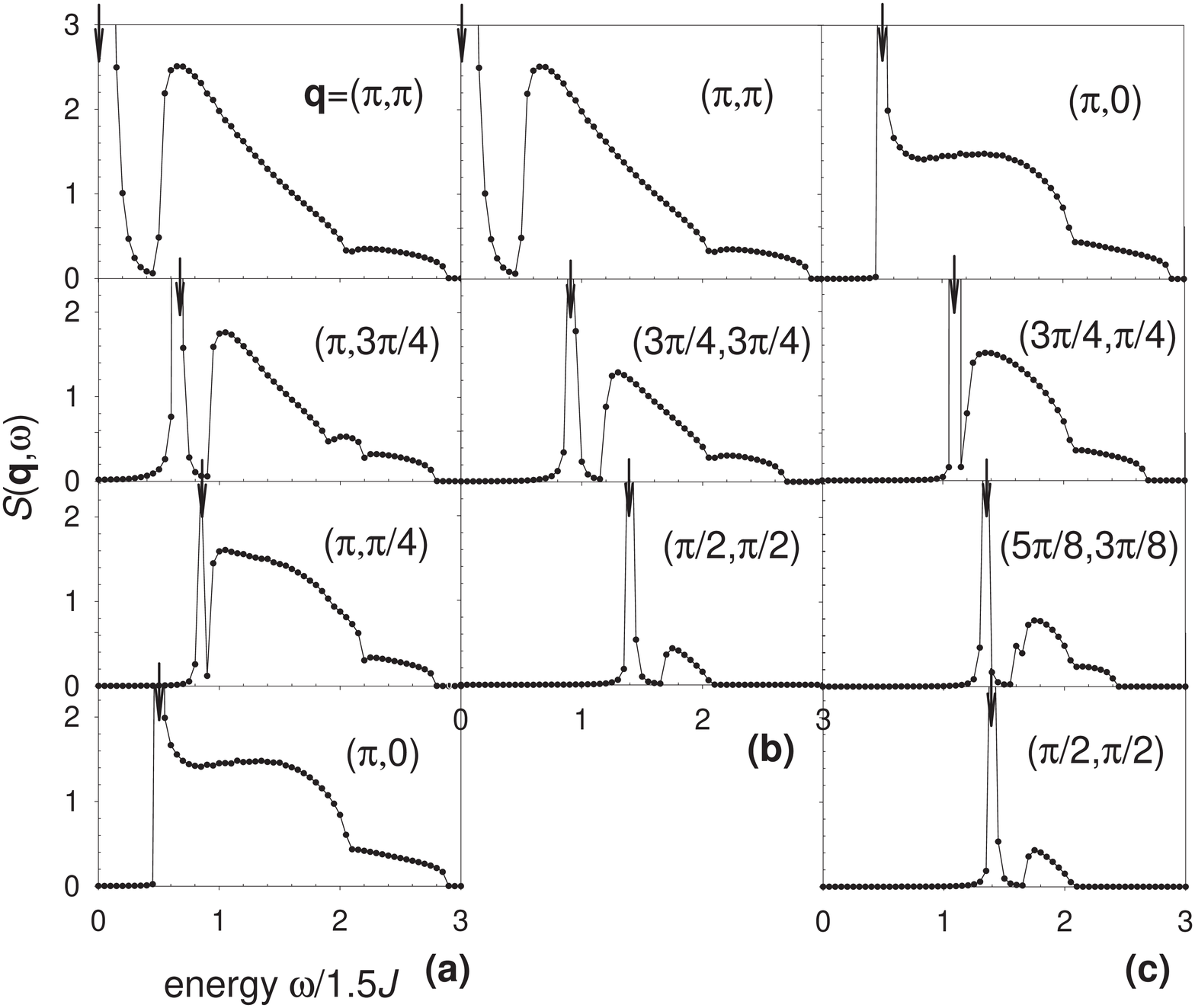}}
\vspace{0.01in}
\caption{Spin susceptibility $S({\bf q}, \omega)$  
as a function of energy $\omega$ for various {\bf q}'s in the 
direction of (a) ($\pi$,$\pi$) to ($\pi$,0), (b) ($\pi$,$\pi$) to
($\pi$/2,$\pi$/2), and (c) ($\pi$,0) to ($\pi$/2,$\pi$/2) along the zone 
boundary. The results are obtained using Eq.(\ref{chi-rpa}) 
with $\delta$=0.001 and 3,000 {\bf q} points in the MBZ.
The energy is in units of 1.5$J$. The small vertical 
arrows mark the positions of the magnon peaks.}
\vspace{0.01in}
\label{fig:suscep}
\end{figure}

\begin{figure}
\epsfxsize=3.0in
\centerline{\epsffile{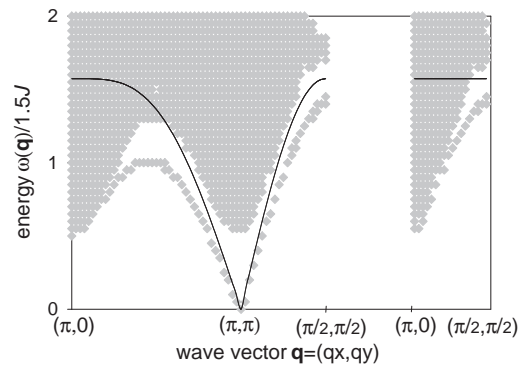}}
\vspace{0.01in}
\caption{Spectrum along high symmetry directions in the MBZ deduced from 
spin susceptibility intensities $S({\bf q},\omega)$. Each (gray) diamond marks 
the $S({\bf q},\omega)$ which has intensity larger than 1.5 (arbitrary 
unit), as chosen here. The low-energy branch represents the magnon dispersion. 
The broad shaded region corresponds to the spinon continuum. The 
solid line represents the 
spin-wave dispersion, $2Z_{c}J \sqrt{1-[(\cos{k_{x}}+\cos{k_{y}})/2]^{2}}$
with $Z_{c} \sim$ 1.18 the renormalization factor.}
\label{fig:spectrum}
\end{figure}

At the magnetic wave vector
${\bf q}$={\bf Q}=$(\pi,\pi)$, $\chi({\bf q},\omega)$ is singular when
$\omega$=0. The spectrum is gapless, which result is guaranteed by the
self-consistency of the RPA. 
Since the gapless mode is a bound state between particle-hole pairs,
it is interpreted as an ``RVB exciton''.
The dispersion of this mode 
can be obtained analytically. Evaluating 
\( \lim_{{\bf q} \rightarrow {\bf Q}} \chi^{+-} \) 
for small $\omega$, we obtain
$\omega({\bf q}) \approx 1.14 J_{eff} |{\bf q}|$.
For arbitrary {\bf q}, Eq.(\ref{chi-rpa}) has to be evaluated
numerically. The results for $S({\bf q},\omega)$ are shown in 
Fig.~\ref{fig:suscep}. For each {\bf q}, we see a strong peak at low
energy.
This magnon (or RVB exciton) peak disperses as ${\bf q}$ varies. 
By tracking the position of the 
intense peaks at low energy, 
we obtain the magnon dispersion for the entire MBZ 
(see Fig.~\ref{fig:spectrum}), recovering Hsu's results \cite{hsu_90}.
Near the zone boundary, the dispersion deviates from that of linear 
spin waves. 

It is important to realize that our approach is much more than just a 
complicated way of describing magnons. 
This is borne out by our explicit
evaluation of $S({\bf q},\omega)$. 
For example, consider $S({\bf q},\omega)$, at ($\pi,\pi$) 
in Fig.~\ref{fig:suscep}.
The intensity drops at higher energy, 
and grows again at $\omega \approx 0.5J_{\rm eff}$. 
The intensity at higher energies is interpreted 
as the manifestation of the particle-hole (spinon)
excitations. Within this interpretation, we can estimate
the characteristic energy at which the continuum becomes visible. 
Consider a spin-flip excitation with momentum {\bf q} as a convolution
of particle-hole excitations. Then, its energy $\omega({\bf q})$ is
given by
$
\omega({\bf q}) = E_{{\bf q-k }} + E_{{\bf k}} 
$,
where we have ignored the residual interaction.
For ${\bf q}$=${\bf Q}$, 
$\omega({\bf Q})$=$2J_{{\rm eff}} \sqrt{\cos k_x^2 +\cos k_y^2 + (m/2)^2}$. 
Its minimum, occurring 
when the two fermions are created at the Dirac 
point $(\pi/2,\pi/2)$, is 
exactly $mJ_{\rm eff}$, as seen in Fig.~\ref{fig:suscep}.
This analysis shows that the
continuum emerging for energies $0.5J_{{\rm eff}}$ and above 
can be interpreted as arising from a 
convolution of free flux fermions or spinons. The line shape is, however, 
determined by the interaction.
Due to the phase space 
available for convolution, the onset of the continuum 
is a strong function 
of momentum. Hence, as seen in Fig.~\ref{fig:suscep}, 
the magnon peak as well as the 
lower boundary 
of the flux-fermion 
continuum shift in energy, as {\bf q} is varied. 
As ${\bf q} \rightarrow (\pi,0)$, the onset of the continuum shifts
towards lower energies. At $(\pi,0)$, there is no
easy way to isolate the 
magnon and spinon contributions \cite{future}.    
In Fig.~\ref{fig:suscep}(b), we display the behavior 
of $S({\bf q},\omega)$ 
along the direction $(\pi,\pi)\rightarrow (\pi/2,\pi/2)$. 
In this case, the 
lower boundary of the fermion continuum shifts towards {\em higher} energies. 
At $(\pi/2,\pi/2)$, we find the onset of the continuum around
$1.7 J_{{\rm eff}}$. As before, this value can be understood by 
considering the
spin-flip excitation with momentum $(\pi/2,\pi/2)$ as a particle-hole
pair. 
A similar analysis as in the previous case yields
a minimum energy, 
$J_{{\rm eff}}(m + 2\sqrt{2+(m/2)^2})/2$ 
at the Dirac point $(\pi/2,\pi/2)$. 
For $m = 0.5$, this turns out to be $1.7J_{{\rm eff}}$, confirming our
interpretation.

From our results for the magnon peak and the fermion continuum, we
obtain the spectrum of excitations over the 
whole MBZ, which is shown in Fig.~\ref{fig:spectrum}. 
Note that by putting $\Phi_{\Box}=0$ in Eq.(\ref{hamiltonian}), results
from the weak coupling approach to the Heisenberg antiferromagnet 
\cite{swz_89} can be obtained. We have verified that the excitation
spectrum, in this case, is distinct from ours \cite{future}.
Thus, our calculations are specific
to the existence of a $\pi$-flux phase \cite{future}.
Our results for small ${\bf q} \simeq {\bf Q}$ 
are consistent with calculations based
on the bosonic RVB scheme \cite{auerbach_88}. At higher energies, it has
been shown \cite{ng_95} that the bosonic RVB state can support vortex
like topological excitations that are fermionic. This agrees with our 
picture of fermionic excitations at high energies.
The shaded region in Fig.~\ref{fig:spectrum} depicts these excitations 
and can, in principle, be observed directly by
INS. At the magnetic wave vector ${\bf Q}$, there 
is a well defined gap between the strong magnon peak and 
the spinon continuum, which may facilitate detection of the continuum 
along this direction.
For typical values of the superexchange coupling $J$ in 
undoped cuprates, this
gap is of the order of 100 meV. 
Recently, Coldea {\em et al.} \cite{coldea_00} have reported 
results from high-energy (0.1-0.5 eV) neutron scattering 
in $La_{2}CuO_{4}$. 
Given the possibility of studying high energy spin excitations 
with enhanced 
resolution, INS may be a good probe for the direct 
observation of spinon 
excitations, should they exist. It is therefore 
extremely interesting to 
examine, both theoretically and experimentally, 
whether our predictions are realized in this measurement. 
Experiments show that as {\bf q} varies from $(\pi,0)$ to
$(\pi/2,\pi/2)$, the spectral weight at high energies decreases. This is
certainly consistent with our calculation.
A detailed comparison incorporating appropriate structure factors 
for neutron scattering is forthcoming \cite{future}. 
 
Let us now turn our attention to the two magnon experiments.
As mentioned earlier, these experiments are characterized by a broad
spectral distribution of intensity and secondary peaks. Both these
features can be explained naturally within our scheme. First, let us
consider the observed widths of the primary peaks in two magnon Raman
scattering. We identify two reasons why the observed linewidths are very
broad. (a) Conventional spin wave theory yields magnons that do not 
disperse along the zone boundary, leading to a singularity in the 
density of magnon states. Consequently, the Raman spectrum is dominated by 
zone boundary magnons. The singularity is, however, 
smoothened by magnon-magnon interactions. 
On the other hand, in the RVB scenario, a broad spectrum is obtained even
with non-interacting magnons. This is because
the low lying mode in Fig.~\ref{fig:spectrum} disperses along the zone 
boundary and there is no singularity in the density of states. Thus, 
the two magnon spectrum is {\em not}
dominated by zone boundary excitations, and is broader than the spectrum
obtained from linear spin wave theory. 
(b) In addition to this intrinsic width of the two magnon spectrum, 
we expect the line shape to be broadened further by the contribution 
from the spinon continuum as is evident from Fig.~\ref{fig:spectrum}.
\begin{figure}
\epsfxsize=1.9in
\centerline{\epsffile{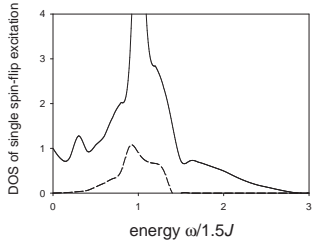}}
\vspace{0.01in}
\caption{Density of states of a single spin-flip excitation. The dashed
line is obtained when only the magnon peak is considered. The solid line
is ${\pi}^{-1} \sum_{\bf q}^{\prime} S({\bf q},\omega)$.}
\vspace{0.01in}
\label{fig:dos}
\end{figure}

To illustrate the above, we calculate the
the density of states (DOS) corresponding to {\em one} 
spin-flip excitation, 
$\sum^{\prime} \delta(\omega-\omega({\bf q}))$. 
In Fig.~\ref{fig:dos}, the dashed line
shows the DOS obtained by considering only the magnon peak. 
To incorporate the entire
excitation spectrum, {\em i.e.}, both the magnon and the spinon
excitations, we also plot (solid line) 
${\pi}^{-1} \sum_{\bf q}^{\prime} S({\bf q},\omega)$.
As seen in the figure, the
presence of spinon excitations leads to a hump at $\omega \sim 2J$, 
followed by a long tail extending to energies $\omega \sim 4J$ 
and beyond. These features are clearly absent when only the magnon peak is
considered. The results indicate that the two
magnon DOS (relevant to Raman experiments) would show a 
primary peak around $\omega \sim 3J$ (from the low lying mode)
and a broad hump at $\omega \sim 4J$ (arising from the continuum).
We expect similar
features in the mid-infrared absorption spectra. In this case, 
the spectrum is broader as the two spin-flip excitation can carry
finite momentum, and the final result involves summing over all such 
momenta \cite{lorenzana_95}. To summarize this discussion, the unconventional dispersion of the
magnon in the RVB scheme leads to broad primary peaks in the optical
spectra and the contribution of the spinon continuum leads to secondary
peaks.

In conclusion, the observation of dispersing magnon
peaks in spin 1/2 Mott insulators does not rule out the existence of
spinons {\em per se}. The magnon can be recovered as an
RVB exciton and the presence of spinons can be
directly observed by a careful analysis of the line shapes obtained
from inelastic neutron scattering. We have obtained, 
within the framework presented in this
paper, the characteristic energy scales at which the
spinon continuum can be observed. Dispersing magnons seen in neutron
scattering and the observation of spectral weight at
high energies in optical experiments, as well as the ubiquitous
secondary peaks, can all be reconciled by our picture. 
While our results do not prove the existence of spinons, they certainly
demonstrate how their presence modifies the spin excitation
spectrum. It is hoped that these
results would spur further experimental investigation of these 
issues, in particular, a careful analysis of the high energy spin
excitation spectrum.
\noindent 

We thank G.~Aeppli, B.~Keimer, P.~A.~Lee, T.~K.~Lee, N.~Nagaosa, T.-K.~Ng, 
and Z.~Y.~Weng for discussions. CMH was supported in Japan by the
Grant-In-Aid for the COE project from Manbusho. Work at Princeton
is supported by NSF Grant DMR-9104873.


\end{multicols}



\begin{thebibliography}{10}
\bibitem{future} We refer to a forthcoming paper for details.
\end{thebibliography}
\end{document}